\documentclass[preprint]{aastex}
\usepackage{txfonts}
\newcommand{\kms}{{\rm km \; s^{-1}}}

\slugcomment{to appear in \emph{The Astronomical Journal}}
\shorttitle{SN~2002cx Late-time Spectroscopy}
\shortauthors{S.~Jha et al.}

\begin{document}
\title{Late-Time Spectroscopy of SN~2002cx: \\
The Prototype of a New Subclass of Type Ia Supernovae}

\author{Saurabh~Jha\altaffilmark{1}, 
David~Branch\altaffilmark{2},
Ryan~Chornock\altaffilmark{1},
Ryan~J.~Foley\altaffilmark{1},\\
Weidong~Li\altaffilmark{1},
Brandon~J.~Swift\altaffilmark{1},
Darrin~Casebeer\altaffilmark{2},
and Alexei~V.~Filippenko\altaffilmark{1}}
\altaffiltext{1}{Department of Astronomy, 601
Campbell Hall, University of California, Berkeley, CA 94720-3411;
\{saurabh, chornock, rfoley, weidong, alex\}@astron.berkeley.edu}
\altaffiltext{2}{Department of Physics and Astronomy, University of
Oklahoma, Norman, OK 73019; \{branch, casebeer\}@nhn.ou.edu}
\setcounter{footnote}{2}

\begin{abstract}
We present Keck optical spectra of SN~2002cx, the most peculiar known
Type Ia supernova (SN Ia), taken 227 and 277 days past maximum
light. Astonishingly, the spectra are not dominated by the forbidden
emission lines of iron that are a hallmark of thermonuclear supernovae
in the nebular phase. Instead, we identify numerous P-Cygni profiles
of \ion{Fe}{2} at very low expansion velocities of $\sim$700 $\kms$,
which are without precedent in SNe Ia. We also report the tentative
identification of low-velocity \ion{O}{1} in these spectra, suggesting
the presence of unburned material near the center of the exploding
white dwarf. SN~2002cx is the prototype of a new subclass of SNe Ia,
with spectral characteristics that may be consistent with recent pure
deflagration models of Chandrasekhar-mass thermonuclear
supernovae. These are distinct from the majority of SNe Ia, for which
an alternative explosion mechanism, such as a delayed detonation, may
be required.
\end{abstract}


\section{Introduction\label{sec:intro}}

The utility of type Ia supernovae (SNe Ia) as excellent cosmological
distance indicators has prompted significant effort into the
observational and theoretical study of this class of cosmic
explosions. While SNe Ia show remarkable homogeneity and quantifiable
heterogeneity, some objects do not fit within the one-parameter
framework that well describes the vast majority of SNe Ia. Perhaps the
most extreme deviant known is SN~2002cx, described by
\citet{Li/etal:2003a}, who presented optical photometry and
spectroscopy from 10 days before to about two months after maximum
light. These data showed SN~2002cx had similar spectral features to
the slowly declining, luminous SN~1991T
\citep{Filippenko/etal:1992,Phillips/etal:1992} near maximum light,
but with significantly lower expansion velocities
\citep{Branch/etal:2004a}. Furthermore, SN~2002cx was subluminous by
$\sim$2 mag in the optical relative to normal SNe Ia, even
though it had a slow late-time decline rate, the opposite of other
subluminous SNe Ia such as SN~1991bg
\citep{Filippenko/etal:1992a,Leibundgut/etal:1993}.

Here we present late-time optical spectra of SN~2002cx taken about
eight months after maximum light. At these epochs, all known SNe Ia
are in their nebular phase, with spectra dominated by forbidden
emission lines of iron and cobalt and an absence of the P-Cygni
profiles of permitted lines that are seen at earlier times
\citep[see][for reviews of SN
spectroscopy]{Filippenko:1997,Branch/Baron/Jeffery:2003}. The low
expansion velocities in SN~2002cx allow for more secure line
identifications than are usually possible in SN spectra because
blending is mitigated. In particular, we anticipated these late-time
observations might resolve the blended forbidden-line emission,
allowing for a check on basic ideas about the structure of SNe Ia
\citep{Kirshner/Oke:1975,Axelrod:1980}, but the results show that the
peculiarities of SN~2002cx persist to these late epochs.

\section{Observations\label{sec:obs}}

Late-time optical spectra of SN~2002cx were taken on 2003 January 7
and 2003 February 27-28 (UT dates are used throughout this paper),
with both beams of the Low Resolution Imaging Spectrometer
\citep[LRIS;][]{Oke/etal:1995} on the Keck I 10-m telescope.  All
observations used the 400/3400 grism on the blue side and the D560
dichroic beamsplitter. The January 7 and February 27 spectra were
taken with the 400/8500 grating on the red side, while the February 28
spectrum used the 1200/7500 grating to better resolve the emission
lines.  A $1\arcsec$ slit was oriented through the nucleus of the host
galaxy for the January observation, but it was aligned with the
parallactic angle \citep{Filippenko:1982} for the February
observations.  Further details of the observations are presented in
Table \ref{tab:obs}. At these epochs SN~2002cx had an approximate
magnitude of $R \simeq$ 21. This is only $\sim$3.5 mag below
peak, an extraordinarily slow decline compared to normal SNe Ia which
fade by $\sim$6 mag over the same time period
\citep[e.g.,][]{Milne/The/Leising:2001}.\footnote{The slow decline is
still consistent with the radioactive decay of $^{56}$Co as the power
source for the late-time light curve, but this question deserves a more
detailed analysis with precise late-time photometry and calculation of
the bolometric luminosity.}

Basic image processing and optimal spectral extraction were performed
in IRAF.\footnote{The Image Reduction and Analysis Facility (IRAF) is
distributed by the National Optical Astronomy Observatories, which are
operated by the Association of Universities for Research in Astronomy,
Inc., under cooperative agreement with the National Science
Foundation.} We used our own IDL procedures to apply a flux
calibration and correct for telluric absorption bands
\citep{Matheson/etal:2000a}. The blue and red halves of the Jan 7
spectrum were scaled and added together over the interval of
wavelength overlap. The blue halves of the Feb 27 and Feb 28 spectra
were taken with the same instrumental setup, so they were added
together before concatenation with the red half of the Feb 27
spectrum.  The red side of the Feb 28 spectrum has higher resolution
than the others, so we have analyzed it separately.

\begin{deluxetable}{lccccccc}
\tablewidth{0pt}
\tabletypesize{\small}
\tablecaption{Spectroscopic Observations of SN~2002cx \label{tab:obs}}
\tablehead{
 \colhead{UT Date} &
 \colhead{Wavelength Range}  &
 \colhead{Resolution} &
 \colhead{Exposure} &
 \colhead{PA} &
 \colhead{Parallactic} &
 \colhead{Airmass} &
 \colhead{Seeing} \\
 \colhead{(Y-M-D)}  &
 \colhead{(\AA)} &
 \colhead{(\AA)} &
 \colhead{(s)} &
 \colhead{($^\circ$)} &
 \colhead{($^\circ$)} &
 \colhead{} &
 \colhead{($\arcsec$)}
}

\startdata
2003-01-07.65 & 3300--9430 & 6 & 2 $\times$ 900  & 153.0 & 132.2 & 1.06 & 1.0\\
2003-02-27.66 & 3160--9420 & 6 & 800             & 66.0  & 67.3  & 1.28 & 0.7\\
2003-02-28.63 & 3160--5770 & 6 & 2 $\times$ 2200 & 240.0 & 242.6 & 1.18 & 0.8\\
2003-02-28.63 & 6260--7540 & 2 & 2 $\times$ 2200 & 240.0 & 242.6 & 1.18 & 0.8\\
\enddata

\end{deluxetable}

\section{Analysis and Results\label{sec:analysis}}

\begin{figure}
\begin{center}
\includegraphics[height=7.2in]{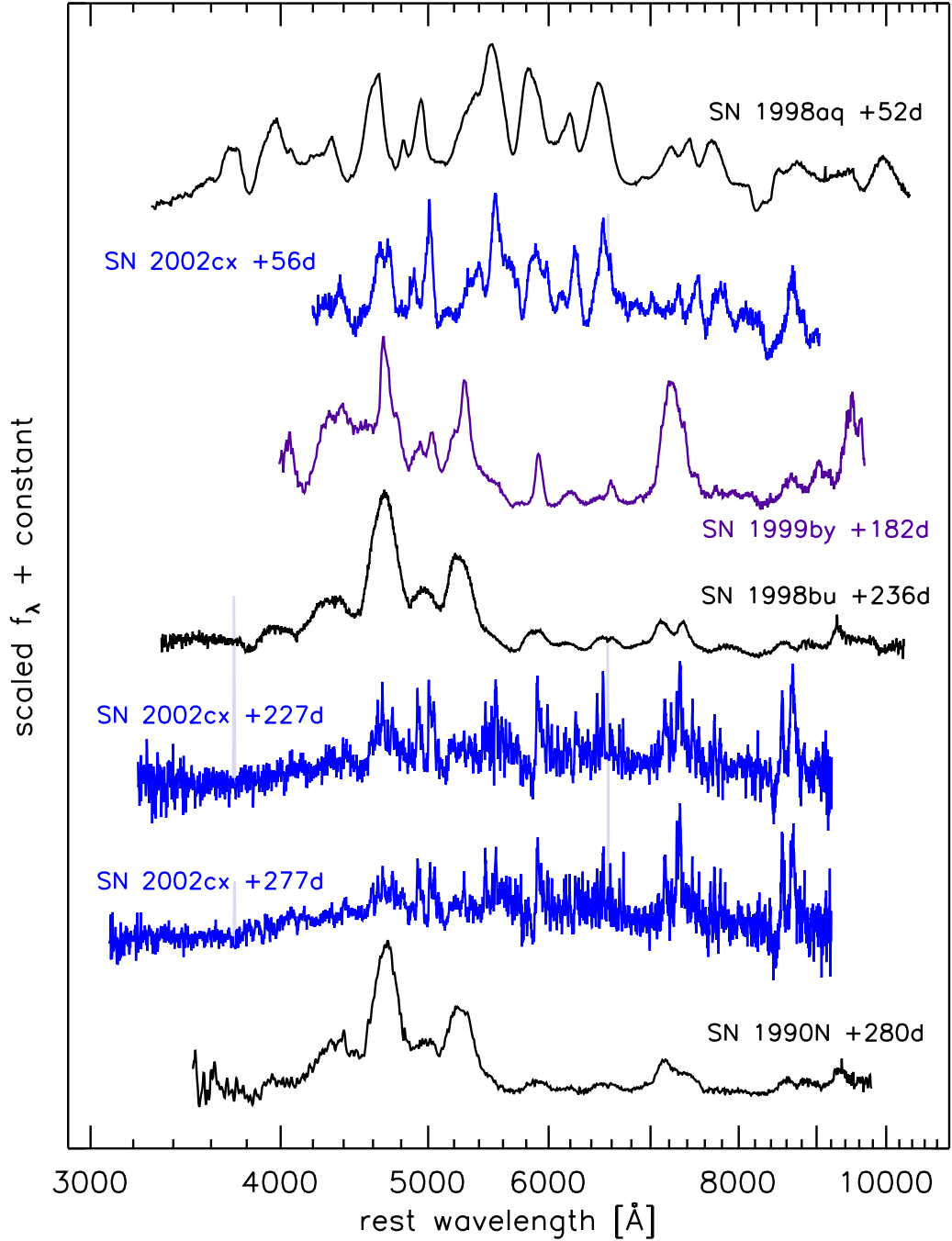}
\end{center}
\vspace{-0.3in}
\caption[Late-time spectra]{\singlespace Optical spectra of SN~2002cx
  (blue) compared with normal SNe Ia at similar epochs (SN~1998aq,
  SN~1998bu, and SN~1990N; black) and the subluminous SN~1999by
  (violet). The spectra have been arbitrarily scaled and
  shifted. Narrow emission lines from an \ion{H}{2} region
  superimposed along the line of sight to SN~2002cx are shown in light
  gray. The epochs listed correspond to supernova rest-frame days past
  $B$ maximum light. The SN~1990N comparison spectrum is from
  \citet{Gomez/Lopez:1998}, while we have observed SN~1998aq on
  1998-Jun-18 and SN~1998bu on 1999-Jan-10 with the Lick Observatory
  3-m Shane telescope (+ KAST), and SN~1999by on 1999-Nov-09 with Keck
  II (+ LRIS).
\label{fig:spectra}}
\end{figure}

The late-time spectra of SN~2002cx, corrected to rest-frame
wavelengths (with $cz = 7184 \; \kms$ for the host galaxy, CGCG
044-035; \citealt{Falco/etal:1999a}), are shown in Figure
\ref{fig:spectra}.\footnote{Throughout this paper, we employ a
logarithmic wavelength axis to facilitate comparison of line widths.}
Narrow, unresolved emission lines of H$\alpha$ and [\ion{O}{2}]
$\lambda$3727 from a superimposed \ion{H}{2} region in the host galaxy
are shown in light gray; they have been excised in subsequent
figures. The January 7 and February 27/28 observations correspond to
227 and 277 days past $B$ maximum light in the SN rest frame
\citep{Li/etal:2003a}. We additionally show the latest SN~2002cx
spectrum from the previous observing season, at an epoch of $+$56
days, presented by \citet{Li/etal:2003a} and analyzed in more detail
by \citet{Branch/etal:2004a}. For comparison we also display spectra
of normal SNe Ia at similar epochs, including SN~1998aq ($+$52 days)
and SN~1998bu ($+$236 days) from our spectral database, and SN~1990N
($+$280 days) from \citet{Gomez/Lopez:1998} via the online SUSPECT
database\footnote{\url{http://suspect.nhn.ou.edu/$\sim$suspect/}}.  We
further show our archive spectrum of the subluminous SN~1999by ($+$182
days) which had an intrinsic peak brightness similar to SN~2002cx, but
with a much faster declining light curve and a very different
early-time spectrum \citep{Garnavich/etal:2004a}.

Despite the clearly peculiar nature of SN~2002cx at early times
\citep{Li/etal:2003a,Branch/etal:2004a}, the day $+56$ spectrum does
not show gross differences compared to the normal SN~1998aq, apart
from lower expansion velocities in the lines. However, the late-time
spectra below appear very different from normal SN Ia counterparts
such as SN~1998bu and SN~1990N, whose flux in the optical is dominated
by broad, blended emission lines of [\ion{Fe}{2}] and [\ion{Fe}{3}]
\citep[e.g.,][]{Ruiz-Lapuente/etal:1995,Mazzali/etal:1998}. Much of
the high-frequency structure in the SN~2002cx observations is due to
real features, and in general, these narrow lines do not correspond to
resolved versions of the features seen in SN~1998bu and SN~1990N.

\begin{figure}
\begin{center}
\includegraphics[scale=0.9,angle=180]{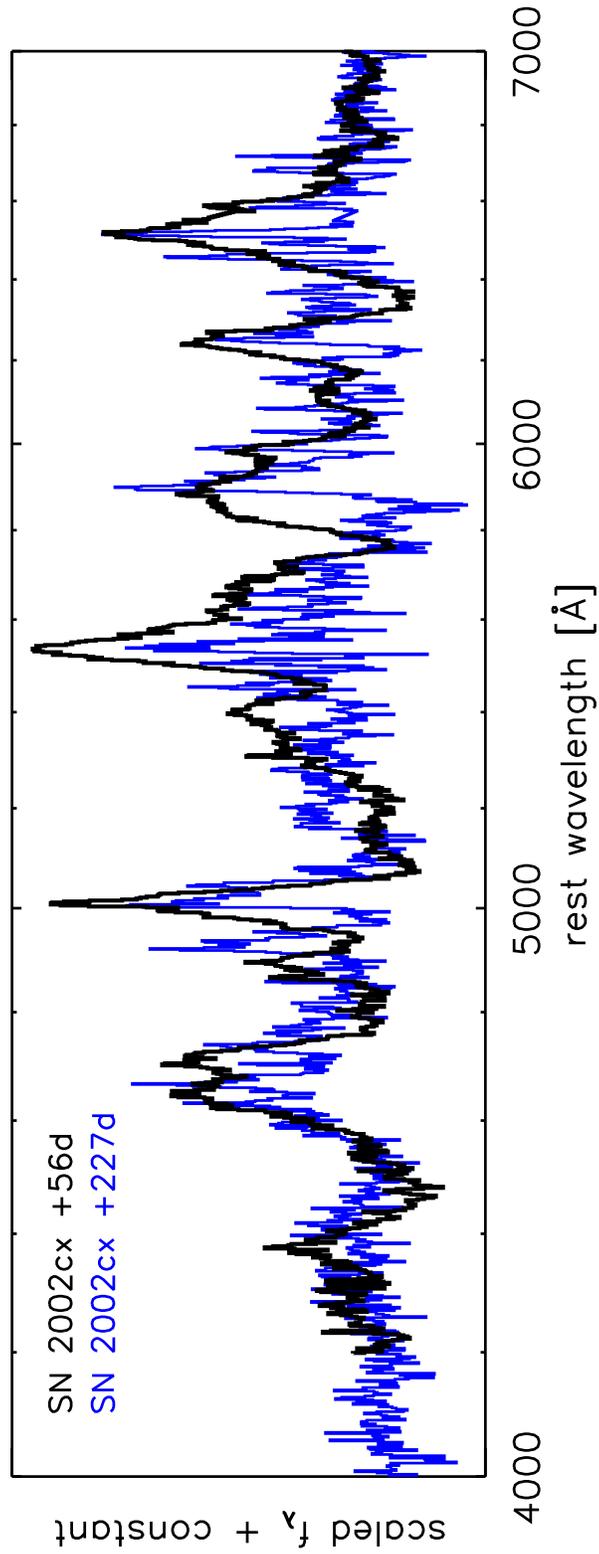}
\end{center}
\caption[Early and Late]{\singlespace Comparison between the day $+$56
  spectrum of SN~2002cx (black) and the day $+$227d spectrum (blue). 
\label{fig:earlylate}}
\end{figure}

The late-time SN~2002cx spectra do show a relatively broad feature
coincident with the strong [\ion{Fe}{3}] 4700 \AA\ feature seen in the
SN~1998bu and SN~1990N spectra. However, unlike normal SNe Ia whose
spectra change dramatically between $\sim$2 months and 8-9 months past
maximum (compare SN~1998aq with SN~1998bu and SN~1990N in Figure
\ref{fig:spectra}), the late-time spectra of SN~2002cx bear a
resemblance to its day $+$56 spectrum, as shown in Figure
\ref{fig:earlylate}. Over this wavelength range, the day $+$227
spectrum merely shows more resolved lines with less blueshifted
absorption. So rather than forbidden iron emission, the broad 4700
\AA\ feature seen at late times in SN~2002cx may be the same species
as in the day $+$56 spectrum, which can be modeled with P-Cygni
profiles of permitted lines of intermediate-mass elements (IMEs),
\ion{Cr}{2}, \ion{Fe}{2}, and \ion{Co}{2} \citep{Branch/etal:2004a}.

To study the relatively narrow features in the $+227$ and $+277$
day spectra, we need a closer look, provided in the four panels of
Figure \ref{fig:detail}.  We also present the higher-resolution red
spectrum from February 28 overplotted on the other spectra (shifted
and scaled to match) in Figure \ref{fig:hires}, showing that the SN
lines are resolved in both the low and high-resolution data (unlike
the narrow emission lines of H$\alpha$, [\ion{N}{2}], and [\ion{S}{2}]
from the \ion{H}{2} region). 

We adopted two approaches to identifying the narrow lines in the
SN~2002cx spectra.  First we attempted to match them with expected
forbidden lines of Co and Fe, but these efforts were largely
unsuccessful. Our calculations assumed simple local thermodynamic
equilibrium (LTE), and though better models of the expected
forbidden-line emission are surely warranted, most of the wavelengths
of the observed narrow lines in SN~2002cx do not correspond well to
strong lines of [\ion{Fe}{1}], [\ion{Fe}{2}], [\ion{Fe}{3}],
[\ion{Co}{1}], [\ion{Co}{2}], or [\ion{Co}{3}].

Our second approach was prompted by the similarity of the late-time
spectra to the day $+56$ spectrum, and we explored the idea that the
same permitted lines seen at earlier times, but now at lower
velocities, could explain the observed late-time spectra.  To this end
we employed the fast synthetic spectrum code Synow
\citep{Branch/etal:2003, Branch/etal:2004a, Branch/etal:2004},
continuing the analysis of \citet{Branch/etal:2004a} to these later
spectra. In Figure \ref{fig:detail} we display a Synow synthetic
spectrum designed to match the day $+227$ spectrum. The model spectrum
(in black) consists of only three ions: \ion{Fe}{2}, \ion{Na}{1}, and
\ion{Ca}{2}, with a photospheric velocity of 650 $\kms$.  For each ion
Table \ref{tab:synow} lists the wavelength of the reference line, the optical
depth at the photosphere, the e--folding velocity for the
exponentially declining optical--depth distribution, and the
excitation temperature.

\begin{figure}
\begin{center}
\includegraphics[scale=0.8,angle=180]{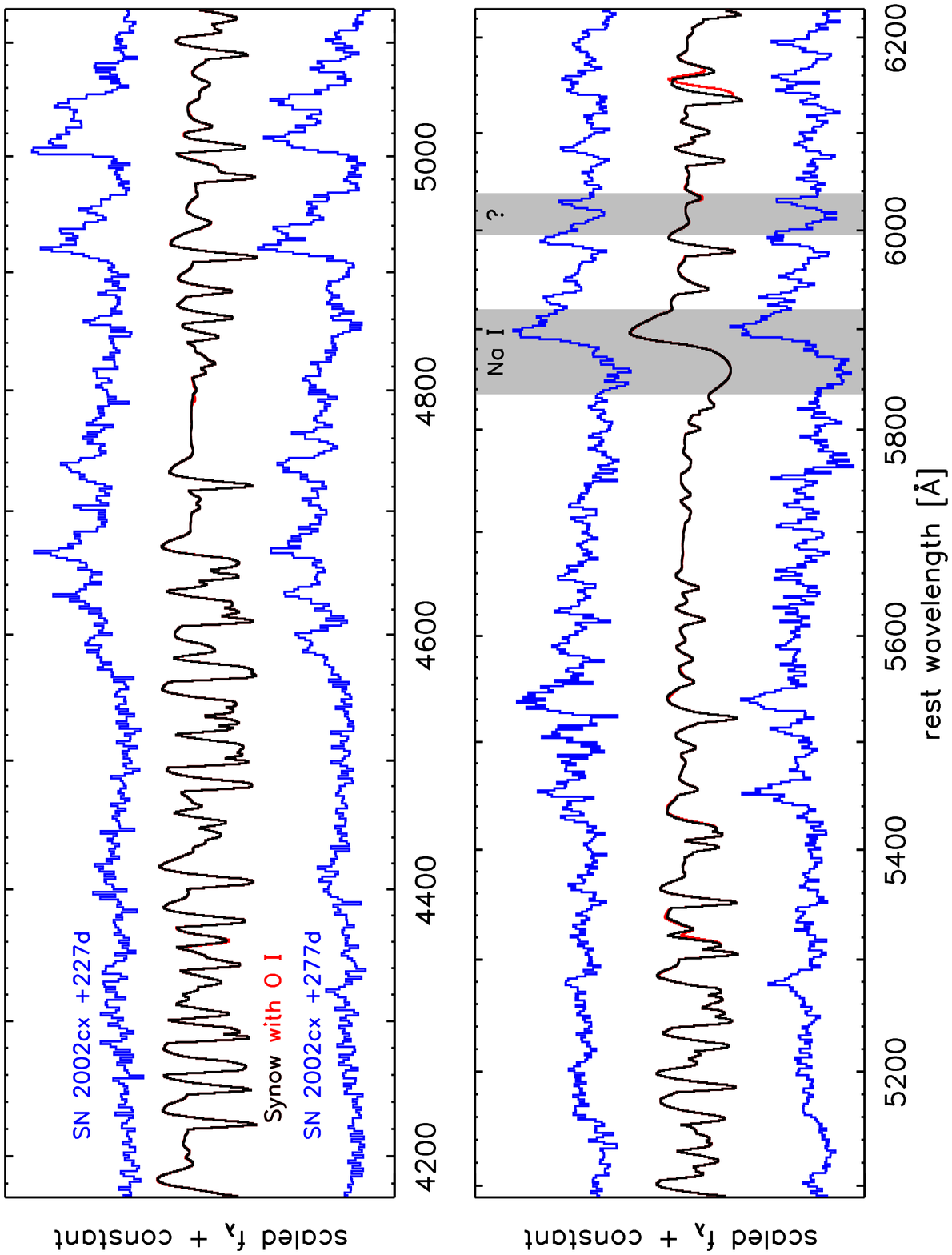}
\end{center}
\caption[Spectra in detail]{\singlespace Expanded views of the
  late-time spectra of SN~2002cx (blue), compared with Synow model
  spectra including \ion{Fe}{2}, \ion{Na}{1}, and \ion{Ca}{2} (black);
  and \ion{O}{1} (red). The shaded regions show line identifications
  (except for \ion{Fe}{2} which are too numerous), and two of the
  strongest unidentified lines.
\label{fig:detail}}
\end{figure}

\addtocounter{figure}{-1}
\begin{figure}
\begin{center}
\includegraphics[scale=0.8,angle=180]{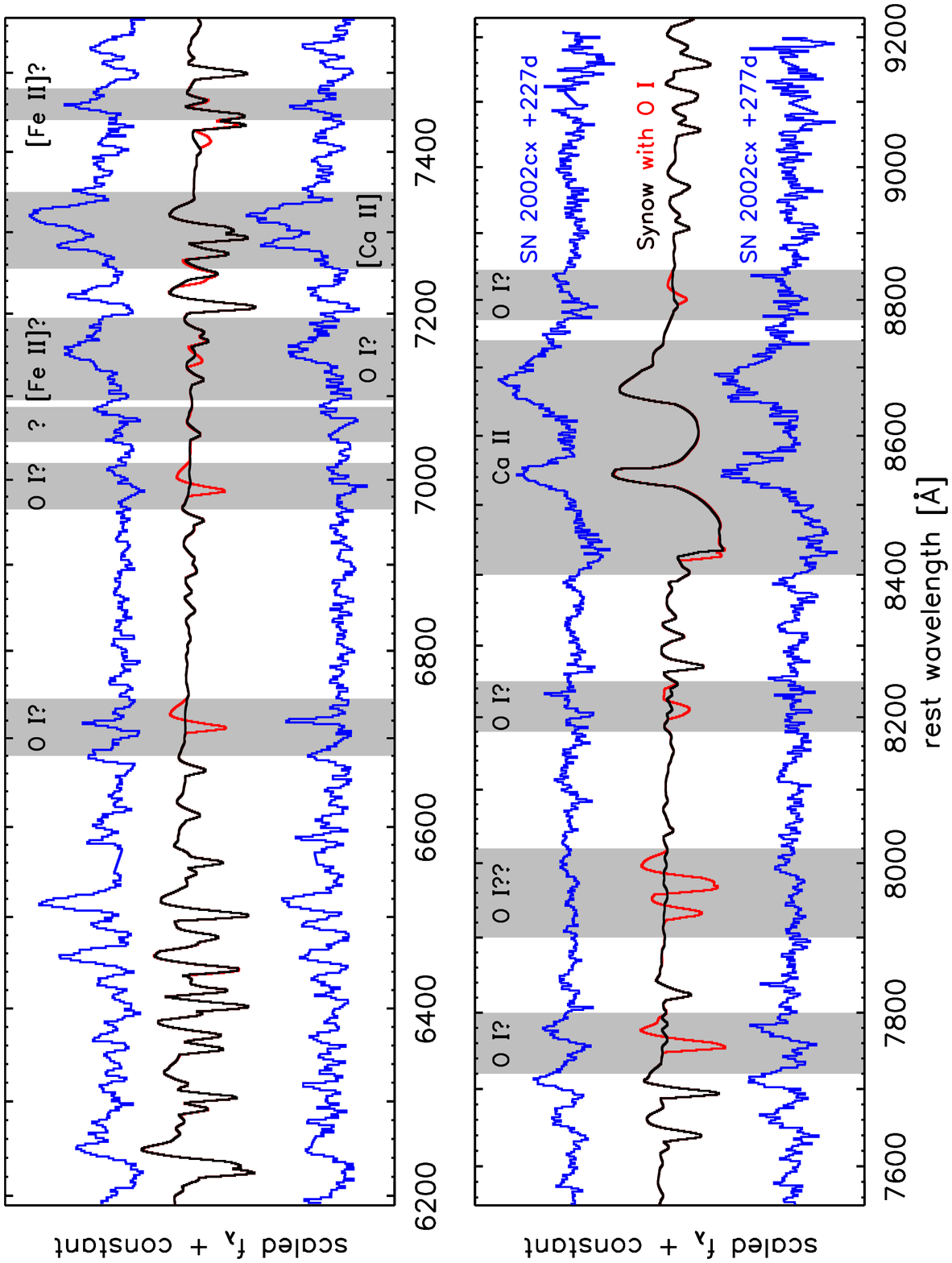}
\end{center}
\caption[Spectra in detail continued]{Continued}
\end{figure}

\begin{deluxetable}{ccccc}
\tablewidth{0pt}
\tablecaption{{\bf Synow} Model Parameters \label{tab:synow}}
\tablehead{
 \colhead{Species} &
 \colhead{$\lambda_{\rm ref}$}  &
 \colhead{$\tau$} &
 \colhead{$v_e$} &
 \colhead{$T_{\rm exc}$} \\
 \colhead{}  &
 \colhead{(\AA)} &
 \colhead{}  &
 \colhead{($\kms$)} &
 \colhead{(K)}
}

\startdata
\ion{Fe}{2} & 5018 & 500 & 1000 & 5000 \\
\ion{Ca}{2} & 3934 & 300 & 3000 & 5000 \\
\ion{Na}{1} & 5890 &   4 & 1000 & 5000 \\
\\
\ion{O}{1} &  7773 & 100 & 1000 & 7000 \\
\enddata

\end{deluxetable}

\begin{figure}
\begin{center}
\includegraphics[scale=0.8,angle=180]{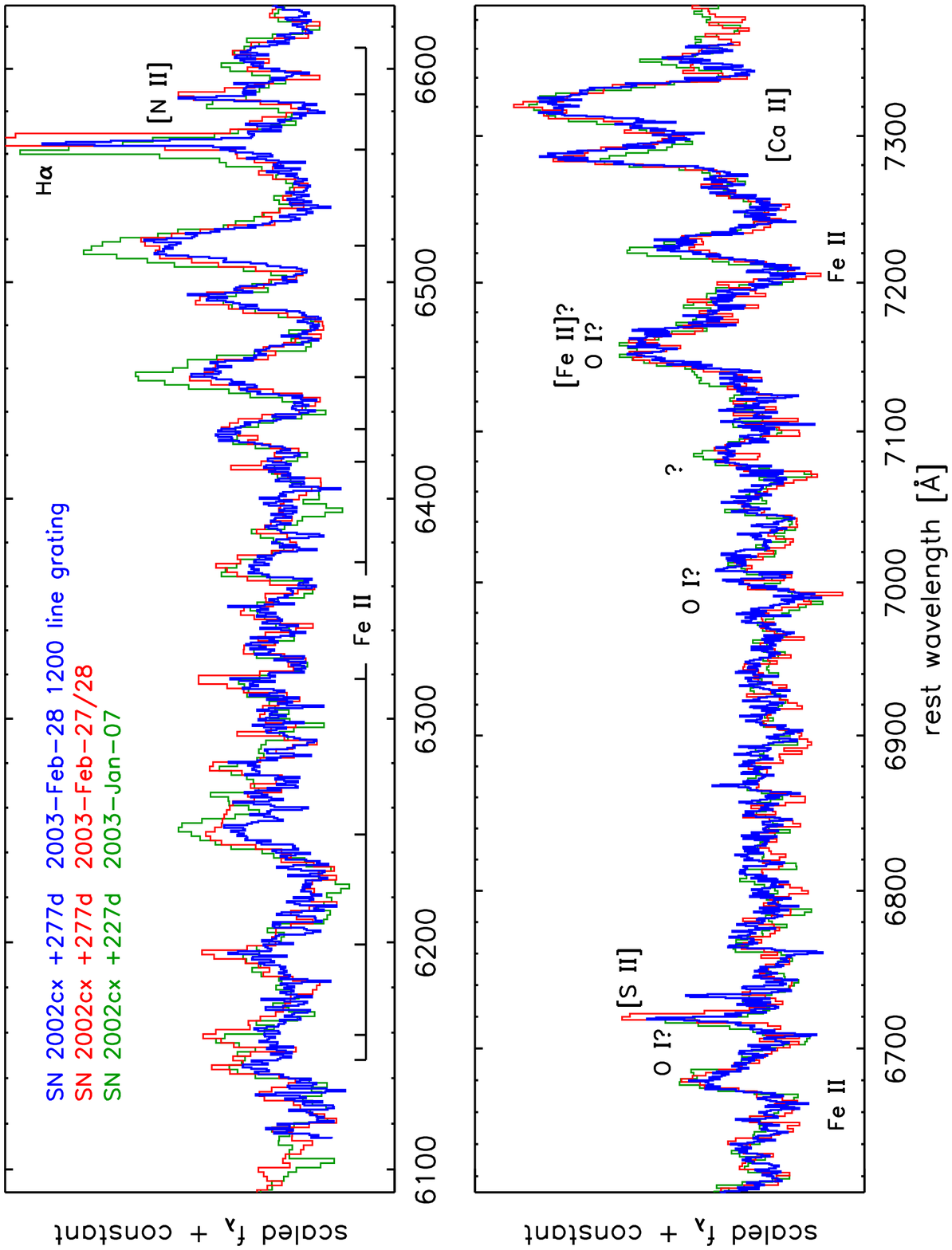}
\end{center}
\caption[Low and high-resolution spectra]{\singlespace Late-time
  spectra of SN~2002cx showing both the low-resolution spectra (red
  and green), as well as the high-resolution spectrum (blue), scaled
  and shifted to match as much as possible. The narrow lines labelled
  H$\alpha$, [\ion{N}{2}], and [\ion{S}{2}] arise from a superimposed
  \ion{H}{2} region.
\label{fig:hires}}
\end{figure}

Below 5800 \AA, the model spectrum consists almost entirely of P-Cygni
profiles of \ion{Fe}{2} resonance-scattering features, which can be
matched (in wavelength, if not always in relative intensity) to the
SN~2002cx spectra starting at approximately 4600 \AA. Blueward of this
the observed spectra may be affected by line blanketing from neutral
species that are not included in the model spectrum. \ion{Fe}{2}
features are also prominent and well-accounted for in the model
between 6200 and 6600 \AA, which can be seen most clearly in the top
panel of Figure \ref{fig:hires}, with the synthetic spectrum lines
marked below. The iron expansion velocities (measured from peak to
trough) are generally consistent over both epochs, but differ slightly
for different lines, with $v \simeq 880 \; \kms$ for \ion{Fe}{2}
$\lambda$5535 and $v \simeq 620 \; \kms$ for \ion{Fe}{2}
$\lambda$6517. The full-width at half-maximum (FWHM) of the emission
part of \ion{Fe}{2} $\lambda$6517 is 650 $\kms$, measured using the
high-resolution spectrum.\footnote{The corresponding value in the
low-resolution data is $\sim$720 $\kms$, showing the effect of
instrumental broadening in the low-resolution data is modest,
amounting to $\sim$300 $\kms$ in quadrature. This is consistent with
the expected broadening based on the instrumental resolution listed in
Table \ref{tab:obs}.} We also note that the broader feature underlying
the narrow H$\alpha$ emission is likely to be \ion{Fe}{2}
$\lambda$6562, rather than a detection of hydrogen associated with the
supernova. The lines seen here are the narrowest ever detected from a
SN Ia; previously the narrowest lines seen were in objects like
SN~1991bg at late times, $v \simeq 2500 \; \kms$
\citep{Mazzali/etal:1997}, but those spectra are quite different from
SN~2002cx at all epochs (as shown by the spectrum of SN~1999by in
Figure \ref{fig:spectra}).

The observed and model spectra also contain features of IMEs, from
\ion{Na}{1} D and \ion{Ca}{2}, which are shown in shaded regions of
Figure \ref{fig:detail}. The profile of \ion{Na}{1} $\lambda$5892 is
broader than the iron lines, with a peak to trough velocity of
$\sim$2200 $\kms$. The blueward absorption extends to at least 4000
$\kms$, and may be present all the way to 8000 $\kms$, but this flux
deficit between 5740 and 5820 \AA\ may be caused by \ion{Fe}{2} lines.
The \ion{Ca}{2} IR triplet is seen in emission, in concert with
[\ion{Ca}{2}] $\lambda\lambda$7291,7324. The forbidden lines have
differing measured FWHM of 450 and 900 $\kms$, but this is likely due
to the strong \ion{Fe}{2} feature seen in the Synow model
spectrum that cuts into the $\lambda$7291 line profile and makes it
seem narrower. The emission components of \ion{Ca}{2} $\lambda$8542
and $\lambda$8662 have FWHM of 1100 and 2000 $\kms$.

The SN~2002cx spectra also show tentative evidence for low-velocity
\ion{O}{1}, which would be the first such detection in a SN Ia. The
Synow model spectrum including \ion{O}{1} in addition to the three
species above is shown in red in Figure \ref{fig:detail}. The
strongest observed features of the putative \ion{O}{1} are
$\lambda$7002 (mostly in absorption, with a velocity at the trough of
500-600 $\kms$, and most clearly visible in Figure \ref{fig:hires})
and $\lambda$7773 (with a similar velocity). The low velocities seen
in SN~2002cx at these epochs significantly constrains line
identifications, and \ion{O}{1} seems the most plausible match for
these two observed features.  Another indication of \ion{O}{1} is
$\lambda$6726 absorption, particularly in the $+227$ day spectrum (the
P-Cygni like appearance of the line is probably mostly due to
contamination of [\ion{S}{2}] $\lambda$6718 from the \ion{H}{2}
region). However, the model predicts strong \ion{O}{1} $\lambda$7948
and $\lambda$7987, which are not clearly observed. Similarly,
identifications of \ion{O}{1} $\lambda$8225 and $\lambda$8820 are
dubious at best.  While unmodeled non-LTE effects may play a role in
the line strengths, on the whole we describe the evidence for
\ion{O}{1} in SN~2002cx at late times as marginal. As we discuss
below, however, its presence at these extremely low velocities would
have significant implications for explosion models.

One possibly important feature is the narrow emission near 7160 \AA\
(with FWHM of 950 $\kms$); this could be \ion{O}{1} $\lambda$7157, but
it would contrast with the other potential \ion{O}{1} lines, which are
mostly in absorption. A more plausible identification of the emission
is [\ion{Fe}{2}] $\lambda$7155, which may be corroborated by the
observed [\ion{Fe}{2}] $\lambda$7453 emission (a weaker transition
from the same upper level). If correct, the presence of these
forbidden iron lines in the red, coupled with their absence in the
blue (specifically, missing strong lines such as [\ion{Fe}{2}]
$\lambda$4289 and $\lambda$5160) attests to how radically different
SN~2002cx is from normal SNe Ia.

It is remarkable that many of the features in the observed SN~2002cx
spectra can be well-matched to a synthetic spectrum with only four
species. Nonetheless, there are still observed lines for which there
is no very plausible identification as \ion{Fe}{2}, \ion{Na}{1},
\ion{Ca}{2}, or \ion{O}{1}. The strongest of these are the P-Cygni
profiles with emission peaks near 6030 and 7080 \AA, denoted with a
question mark in Figures \ref{fig:detail} and \ref{fig:hires}, and
there are many more differences in detail between the observed and
Synow model spectra.

While more sophisticated modeling of these late-time spectra is
clearly necessary, we can make some initial attempts at discerning
physical conditions in SN~2002cx. The mere presence of P-Cygni
profiles of resonance lines implies the ejecta must still be producing
a continuum or pseudo-continuum, at odds with the optically thin
models of normal SNe Ia at these epochs. The best-guess Synow model
starts with a relatively hot thermal continuum ($\sim$15,000 K) in
attempting to match the observed spectra, but this overpredicts the
flux below 4600 \AA\ and above 7500 \AA. Moreover, the observations
likely suffer from contamination in the blue from an underlying star
cluster, further confounding attempts at identifying a continuum from
SN~2002cx.

The observed forbidden lines also provide temperature and density
diagnostics: for example, the [\ion{Ca}{2}] emission strength is on
the same order as the IR triplet emission, typical of gas with an
electron number density $n \simeq 10^{9.5}$ cm$^{-3}$
\citep{Ferland/Persson:1989,Fransson/Chevalier:1989}. Such a high
density could explain the lack of other forbidden lines that
might otherwise be expected, such as [\ion{O}{1}] $\lambda$6300. This
high density would also imply a high mass at low velocity: a naive
calculation based on a constant velocity expansion of 700 $\kms$ over
$\sim$300 days after the explosion yields a mass of 0.4 to 1.0
$M_{\sun}$, assuming a filling factor of 0.3 and a mean atomic weight
$A$ = 20 to 50. However, this estimate is uncertain because of
clumping or otherwise more complicated emission regions. The data do
suggest a gradient in conditions; for example, the IR triplet lines
are broader than the forbidden emission. Additionally, if the feature
near 7160 \AA\ is [\ion{Fe}{2}] $\lambda$7155, the presence of both
forbidden and permitted iron lines suggests a lower density, $n \simeq
10^{7}$ cm$^{-3}$ \citep[e.g.,][]{Verner/etal:2002}. In this case, the
lack of [\ion{Fe}{2}] emission in the blue may indicate a relatively
low excitation temperature $T \lesssim$ 3000 K, which would suppress
otherwise strong lines like $\lambda$4289 and $\lambda$5160 that
originate from a higher energy level than $\lambda$7155. We reiterate,
however, that a more detailed analysis including non-LTE effects is
required for a definitive interpretation of the spectra.

\section{Discussion\label{sec:disc}}

The standard model for a thermonuclear supernova is the explosion of a
C/O white dwarf, but the details of this process that lead to the
observed general homogeneity and correlated differences (in
luminosity, light-curve shape, and spectral features) among SNe Ia are
not fully understood. Open questions at the forefront of current
research include \citep[for a review,
see][]{Hillebrandt/Niemeyer:2000} the nature of the nuclear burning
front, and in particular, whether it remains a subsonic deflagration
or if a transition to a detonation is required (pure detonations are
ruled out due to overproduction of iron-peak elements at the expense
of IMEs); the site and number of ignition points; and even more basic
unsettled issues such as the mass of the white dwarf (Chandrasekhar or
sub-Chandra) and the nature of the companion star (single degenerate
vs. double degenerate systems).

Peculiar objects like SN~2002cx hold the promise of defining the
fringes of the distribution, and from their differences, may
indirectly help to explain the properties of normal SNe Ia as
well. Under the assumption that the explosion leads to homologous
expansion, so that expansion velocities can be used as a proxy for
radius, the important characteristics of SN~2002cx, based on its light
curve \citep{Li/etal:2003a}, early-time spectra
\citep{Branch/etal:2004a}, and late-time spectra are as follows:

1. Low expansion velocities at all epochs, with iron features at
   $\sim$7000 $\kms$ near maximum, $\sim$2000 $\kms$ in the day $+56$
   spectrum, and $\sim$700 $\kms$ in the late-time spectra. This
   implies the presence of fully burned (i.e., to the iron peak)
   material in all layers, and low kinetic energy in the ejecta.

2. Very low peak luminosity (similar to SN~1991bg-like objects),
   implying the production of only a small amount of $^{56}$Ni that
   powers the light curve \citep{Arnett:1982,Pinto/Eastman:2000}.

3. Moderate post-maximum decline rate in $B$ ($\Delta m_{15}(B) = 1.29
   \pm 0.11$ mag), but a much slower decline and broader light curve
   in $R$, and a very slow late-time decline compared to all other SNe
   Ia.

4. SN~1991T-like \ion{Fe}{3} features dominating the early time
   spectrum, implying the outer layers were relatively hot
   \citep{Nugent/etal:1995a}.

5. Intermediate-mass elements present in all layers: Na, Si, S, and Ca
   seen at early times ($v \simeq 7000 \; \kms$); Na (700--4000
   $\kms$) and Ca (700--2000 $\kms$) seen at late times.

6. Permitted \ion{Fe}{2} lines and continuum or pseudo-continuum
   flux at late times (when normal SNe Ia are in the nebular phase),
   suggesting a relatively high density and large mass at low
   velocity.

7. \emph{Possible} presence of \ion{O}{1} at low velocity (500--1000
   $\kms$) in the late-time spectra, implying \emph{unburned} material
   in the inner layers.
   
\citet{Li/etal:2003a} concluded that no published one-dimensional SN
Ia explosion model could reproduce all of the observed properties of
SN~2002cx. Indeed, the late-time spectra further suggest that
iron-peak and intermediate-mass elements are mixed throughout the
ejecta, with unburned material possibly detected close to the
center. This all points away from a fully stratified radial structure,
and suggests three-dimensional models may be required.
\citet{Branch/etal:2004a} noted that recent three-dimensional models
of pure deflagrations in a Chandrasekhar-mass C/O white dwarf may hold
promise for explaining observations of SN~2002cx.

The results of \citet*{Reinecke/Hillebrandt/Niemeyer:2002a} and
\citet{Gamezo/etal:2003} suggest that the convoluted structure of the
turbulent flame in pure deflagrations generically predict ejecta of
mixed composition, and the presence of partially burned (IMEs) and
unburned material (C and O) near the white dwarf center. This latter
point is generally considered a shortcoming of the model since
low-velocity oxygen or carbon is not seen in typical SNe Ia. It may be
hidden at early times \citep*{Baron/Lentz/Hauschildt:2003}, but if
present, should be detectable at late times in nebular spectra of
normal SNe Ia \citep{Kozma/etal:2005}.
\citet*{Gamezo/Khokhlov/Oran:2004,Gamezo/Khokhlov/Oran:2005} show that
artificially introducing a transition to a delayed detonation into
their deflagration model causes a new burning front to sweep through
the unburned and partially burned central material, leaving only
iron-peak elements in the core, and bringing the model in accord with
observations.

In this framework, the strong evidence for partially burned material
(intermediate mass elements) and tentative evidence for unburned
material (oxygen) near the center of SN~2002cx can naturally be
explained by a pure deflagration (or otherwise stopping the detonation
wave from reaching the center). Furthermore,
\citet{Gamezo/Khokhlov/Oran:2005} find their deflagration model
produces about 40\% of the total energy and $^{56}$Ni mass of their
delayed detonation model, which could explain both the low ejecta
velocities and the subluminosity of SN~2002cx relative to normal
SNe~Ia. However, the SN~2002cx spectra presented here do not match the
deflagration model spectra of \citet{Kozma/etal:2005}, which predict a
forbidden-line dominated spectrum, including very strong [\ion{O}{1}]
$\lambda$6300. While the lack of this emission in the observed spectra
is qualitatively consistent with the high density inferred from the
forbidden-line density diagnostics, it may be a difficult challenge
for deflagration models to maintain such high densities and large mass
at low velocities. Moreover, it will be important to explore whether
the other peculiar properties of SN~2002cx (its early-time
SN~1991T-like spectrum and its distinct light curve) are natural
predictions of a pure-deflagration model.  If future deflagration
models prove successful at explaining SN~2002cx, this could help
establish the tantalizing possibility that normal SNe~Ia
\emph{require} a transition to a delayed detonation or something
similar, lest they all look like SN~2002cx.

\begin{figure}
\begin{center}
\includegraphics[width=6in]{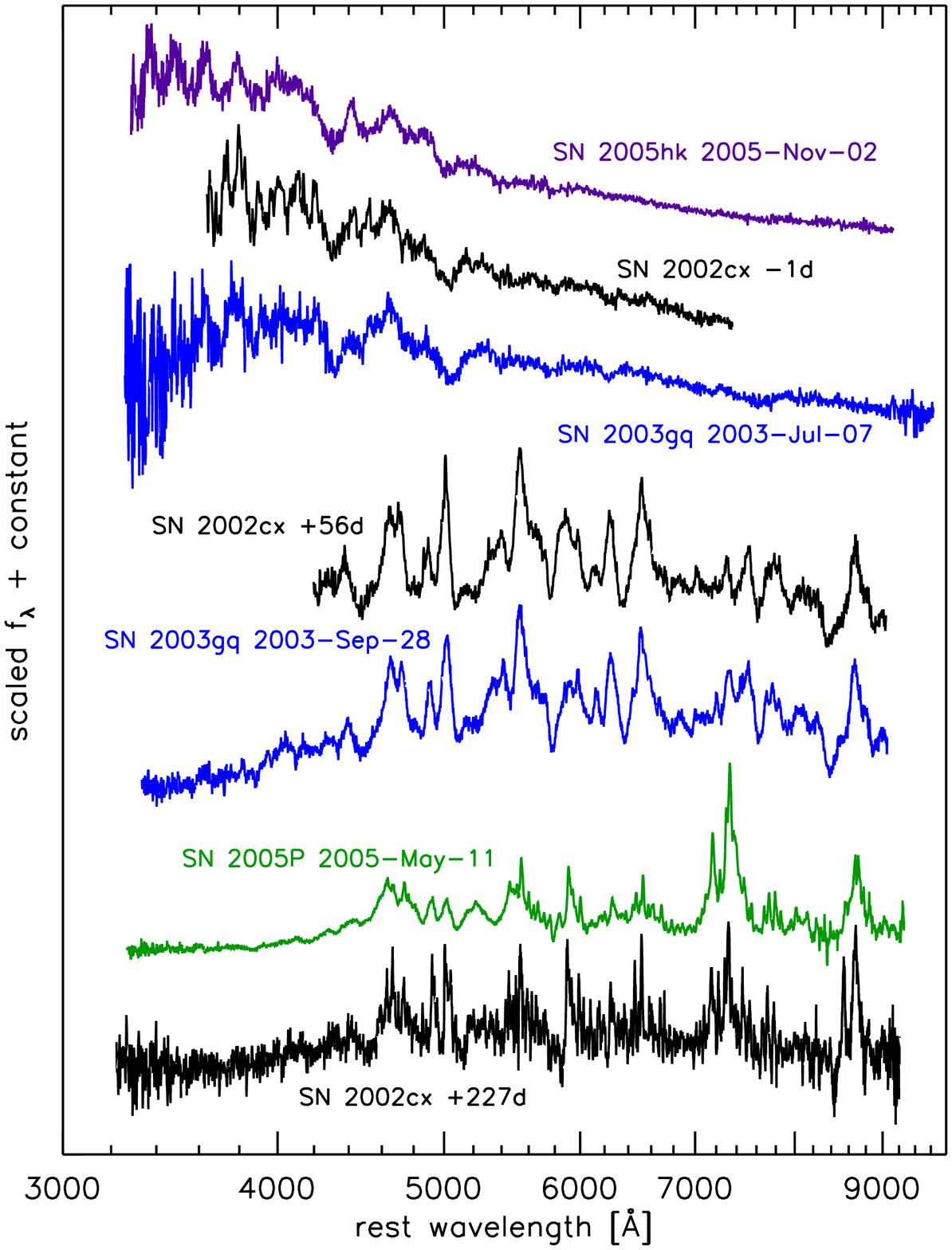}
\end{center}
\caption[Late-time spectra]{\singlespace Spectra of SN~2005hk
  (violet), SN~2003gq (blue), and SN~2005P (green), compared to
  SN~2002cx \citep[][and this paper]{Li/etal:2003a}. The 2005-Nov-02
  SN~2005hk and 2003-Jul-07 SN~2003gq spectra were obtained with the
  Lick Observatory 3-m Shane telescope (+ KAST), while the 2003-Sep-28
  SN~2003gq and the 2005-May-11 SN~2005P spectra were taken with Keck
  I (+ LRIS).
\label{fig:others}}
\end{figure}

Furthermore, recent observations have shown that SN~2002cx is no
longer a unique object. In Figure \ref{fig:others}, we present a
comparison of SN~2002cx with three other SNe Ia in our spectral
database, showing remarkable similarities.  SN~2003gq
\citep*{Puckett/Langoussis:2003,Graham/Weisz/Li:2003} was classified
as a 1991T-like SN Ia from its early-time spectrum (2003 July 7) shown
in Figure \ref{fig:others}, with \ion{Fe}{3} features and weak
\ion{Ca}{2} and \ion{Si}{2} \citep*{Filippenko/Foley/Desroches:2003},
but it developed SN~2002cx-like features at later times
\citep{Filippenko/Chornock:2003}; the spectrum of SN~2003gq on 2003
September 28 is nearly identical (amazingly so) to the $+56$ day
spectrum of SN~2002cx. In addition, a spectrum of SN~2005P
\citep{Burket/Li:2005}, obtained on 2005 January 23 by P.~Wood and
B.~Schmidt (priv.~communication), was identified by one of us (RC) as
an analogue of SN~2002cx. We obtained a spectrum of SN~2005P on 2005
May 11, shown in Figure \ref{fig:others}, which seems to be at an
intermediate epoch between the day $+56$ and $+227$ spectra of
SN~2002cx. A third object, SN~2005cc \citep{Puckett/Langoussis:2005},
has also been shown to have SN~2002cx-like features, and may have been
observed a week before maximum light
\citep{Antilogus/etal:2005}. Finally, SN~2005hk
\citep{Burket/Li:2005a,Barentine/etal:2005} was classified as a SN
1991T-like object based on an spectrum taken a week before maximum
light \citep{Serduke/Wong/Filippenko:2005}. This spectrum is also
shown in Figure \ref{fig:others}; subsequent photometry and
spectroscopy of SN~2005hk reveal it to be similar to SN~2002cx
(\citealt{Chornock/etal:2006}; Phillips et al.~2006, in preparation).

SN~2002cx is thus the prototype of a new class of peculiar SNe Ia.
The striking spectral homogeneity within this subclass suggests these
objects should share a common explanation. As discussed by
\citet{Li/etal:2003}, SN~2002cx-like objects are certainly SNe Ia:
their maximum-light spectra show the same features as SNe Ia like
SN~1991T, and these similarities persist for a significant time
(compare, for example, SN~2002cx and the normal SN~Ia 1998aq at two
months past maximum in Figure \ref{fig:spectra}). Though these
features are at lower velocities in the SN~2002cx subclass, this in
itself is not a compelling reason to exclude them as SNe Ia as
traditionally (spectroscopically) defined. Nonetheless, the large
differences between this subclass and other SNe Ia may require a
radically different model for these objects, perhaps much more
significant than just a pure deflagration. Still more speculative
models, such as ``failed'' SNe Ia in which the white dwarf is not
completely disrupted \citep*[e.g.,][and references
therein]{Livne/Asida/Hoflich:2005}, or perhaps even core-collapse
scenarios, may be necessary. Note, however, that normal
stripped-envelope supernovae exhibit very strong [\ion{O}{1}]
$\lambda$6300 emission at late times; hence, a successful
core-collapse model for SN~2002cx-like objects would have to differ
substantially from typical SNe Ib/Ic.

Additional observations will test these speculations.  Because
SN~2005P (host NGC 5468; $cz =$ 2842 $\kms$;
\citealt{Koribalski/etal:2004}), SN~2005cc (NGC 5383; $cz =$ 2270
$\kms$; \citealt{vanDriel/etal:2001}), and SN~2005hk (UGC 272; $cz =$
3912 $\kms$; \citealt{Abazajian/etal:2003}) are relatively nearby,
they provide promising opportunities to study members of the SN~2002cx
subclass at even later epochs and over a wider wavelength region than
the data presented here. It will be especially important to confirm
the low-velocity unburned material suggested by the SN~2002cx spectra,
through a definitive identification of oxygen or carbon lines at late
times.  The properties of this subclass are just beginning to be
explored, but already their spectral similarities will significantly
constrain models; for instance a pure deflagration or other successful
model of this subclass should generically produce a hot outer layer
with \ion{Fe}{3} features in spectra near maximum light. In addition,
all five of these SN~2002cx-like objects occurred in blue,
star-forming spiral galaxies, further connecting them to SN~1991T-like
objects \citep{Hamuy/etal:2000}, and making the subluminosity of
SN~2002cx and its analogues all the more puzzling. The existence of
this class may also disfavor tuned or potentially ``rare'' models,
depending on the homogeneity among these objects and their frequency
relative to normal SNe Ia. For example, the identification of several
such SNe (and their location in late-type hosts) disfavors a suggested
model of \citet{Kasen/etal:2004}, who hypothesized that SN~2002cx was
an intrinsically subluminous explosion like SN~1991bg, viewed from a
special geometry through a hole in the ejecta to produce a hotter
spectrum at lower velocities. While the application of conclusions
based on SN~2002cx-like objects may not directly answer all of the
outstanding questions regarding the vast majority of normal SNe Ia,
further observations of these examples on the fringe are certain to be
illuminating.

\acknowledgments

We thank Frank Serduke for help with data reduction, and we appreciate
staff assistance at the Keck and Lick Observatories. SJ gratefully
acknowledges support via a Miller Research Fellowship at UC
Berkeley. Supernova research at UC Berkeley is supported by NSF grant
AST-0307894 to AVF, and at Oklahoma by NSF grant AST-0506028 to DB.
This research has made use of the NASA/IPAC Extragalactic Database
(NED) which is operated by the Jet Propulsion Laboratory, Caltech,
under contract with the National Aeronautics and Space
Administration. Data presented herein were obtained at the W. M. Keck
Observatory, which is operated as a scientific partnership among the
the University of California, Caltech, and NASA, made possible by the
generous financial support of the W. M. Keck Foundation. The authors
recognize and acknowledge the very significant cultural role and
reverence that the summit of Mauna Kea has always had within the
indigenous Hawaiian community, and we are most privileged to have the
opportunity to explore the Universe from this mountain.

\bibliography{refs-sn02cx}

\begin{thebibliography}{53}
\expandafter\ifx\csname natexlab\endcsname\relax\def\natexlab#1{#1}\fi

\bibitem[{{Abazajian} {et~al.}(2003){Abazajian}, {Adelman-McCarthy},
  {Ag{\"u}eros}, {Allam}, {Anderson}, {Annis}, {Bahcall}, {Baldry}, {Bastian},
  {Berlind}, {Bernardi}, {Blanton}, {Blythe}, {Bochanski}, {Boroski},
  {Brewington}, {Briggs}, {Brinkmann}, {Brunner}, {Budav{\'a}ri}, {Carey},
  {Carr}, {Castander}, {Chiu}, {Collinge}, {Connolly}, {Covey}, {Csabai},
  {Dalcanton}, {Dodelson}, {Doi}, {Dong}, {Eisenstein}, {Evans}, {Fan},
  {Feldman}, {Finkbeiner}, {Friedman}, {Frieman}, {Fukugita}, {Gal},
  {Gillespie}, {Glazebrook}, {Gonzalez}, {Gray}, {Grebel}, {Grodnicki}, {Gunn},
  {Gurbani}, {Hall}, {Hao}, {Harbeck}, {Harris}, {Harris}, {Harvanek},
  {Hawley}, {Heckman}, {Helmboldt}, {Hendry}, {Hennessy}, {Hindsley}, {Hogg},
  {Holmgren}, {Holtzman}, {Homer}, {Hui}, {Ichikawa}, {Ichikawa}, {Inkmann},
  {Ivezi{\'c}}, {Jester}, {Johnston}, {Jordan}, {Jordan}, {Jorgensen},
  {Juri{\'c}}, {Kauffmann}, {Kent}, {Kleinman}, {Knapp}, {Kniazev}, {Kron},
  {Krzesi{\'n}ski}, {Kunszt}, {Kuropatkin}, {Lamb}, {Lampeitl}, {Laubscher},
  {Lee}, {Leger}, {Li}, {Lidz}, {Lin}, {Loh}, {Long}, {Loveday}, {Lupton},
  {Malik}, {Margon}, {McGehee}, {McKay}, {Meiksin}, {Miknaitis}, {Moorthy},
  {Munn}, {Murphy}, {Nakajima}, {Narayanan}, {Nash}, {Neilsen}, {Newberg},
  {Newman}, {Nichol}, {Nicinski}, {Nieto-Santisteban}, {Nitta}, {Odenkirchen},
  {Okamura}, {Ostriker}, {Owen}, {Padmanabhan}, {Peoples}, {Pier}, {Pindor},
  {Pope}, {Quinn}, {Rafikov}, {Raymond}, {Richards}, {Richmond}, {Rix},
  {Rockosi}, {Schaye}, {Schlegel}, {Schneider}, {Schroeder}, {Scranton},
  {Sekiguchi}, {Seljak}, {Sergey}, {Sesar}, {Sheldon}, {Shimasaku}, {Siegmund},
  {Silvestri}, {Sinisgalli}, {Sirko}, {Smith}, {Smol{\v c}i{\'c}}, {Snedden},
  {Stebbins}, {Steinhardt}, {Stinson}, {Stoughton}, {Strateva}, {Strauss},
  {SubbaRao}, {Szalay}, {Szapudi}, {Szkody}, {Tasca}, {Tegmark}, {Thakar},
  {Tremonti}, {Tucker}, {Uomoto}, {Vanden Berk}, {Vandenberg}, {Vogeley},
  {Voges}, {Vogt}, {Walkowicz}, {Weinberg}, {West}, {White}, {Wilhite},
  {Willman}, {Xu}, {Yanny}, {Yarger}, {Yasuda}, {Yip}, {Yocum}, {York},
  {Zakamska}, {Zehavi}, {Zheng}, {Zibetti}, \& {Zucker}}]{Abazajian/etal:2003}
{Abazajian}, K. {et~al.} 2003, \aj, 126, 2081

\bibitem[{{Antilogus} {et~al.}(2005){Antilogus}, {Garavini}, {Gilles}, {Pain},
  {Aldering}, {Bailey}, {Lee}, {Loken}, {Nugent}, {Perlmutter}, {Scalzo},
  {Thomas}, {Wang}, {Weaver}, {Bonnaud}, {Pecontal}, {Blanc}, {Bongard},
  {Copin}, {Gangler}, {Sauge}, {Smadja}, {Kessler}, {Baltay}, {Rabinowitz}, \&
  {Bauer}}]{Antilogus/etal:2005}
{Antilogus}, P. {et~al.} 2005, The Astronomer's Telegram, 502, 1

\bibitem[{{Arnett}(1982)}]{Arnett:1982}
{Arnett}, W.~D. 1982, \apj, 253, 785

\bibitem[{{Axelrod}(1980)}]{Axelrod:1980}
{Axelrod}, T.~S. 1980, Ph.D.~Thesis, University of California, Santa Cruz

\bibitem[{{Barentine} {et~al.}(2005){Barentine}, {Bassett}, {Becker}, {Bender},
  {Bremer}, {Brewington}, {Dejongh}, {Dembicky}, {Depoy}, {Dilday}, {Doi},
  {Edge}, {Elson}, {Frieman}, {Garnavich}, {Goobar}, {Harvanek}, {Gueth},
  {Holtzman}, {Hopp}, {Kollatschny}, {Konishi}, {Krzesinski}, {Lamenti},
  {Lampeitl}, {Kessler}, {Ketzeback}, {Long}, {Marriner}, {Marshall},
  {McMillan}, {Mendez}, {Miknaitis}, {Morokuma}, {Nichol}, {Pan}, {Prieto},
  {Richmond}, {Riess}, {Romani}, {Romer}, {Ruiz-Lapuente}, {Sako}, {Schneider},
  {Smith}, {Snedden}, {Subbarao}, {Takanashi}, {Tokita}, {van der Heyden},
  {Wheeler}, \& {Yasuda}}]{Barentine/etal:2005}
{Barentine}, J. {et~al.} 2005, Central Bureau Electronic Telegrams, 268, 1

\bibitem[{{Baron} {et~al.}(2003){Baron}, {Lentz}, \&
  {Hauschildt}}]{Baron/Lentz/Hauschildt:2003}
{Baron}, E., {Lentz}, E.~J., \& {Hauschildt}, P.~H. 2003, \apjl, 588, L29

\bibitem[{{Branch} {et~al.}(2003{\natexlab{a}}){Branch}, {Baron}, \&
  {Jeffery}}]{Branch/Baron/Jeffery:2003}
{Branch}, D., {Baron}, E., \& {Jeffery}, D.~J. 2003{\natexlab{a}}, LNP
  Vol.~598: Supernovae and Gamma-Ray Bursters, 598, 47

\bibitem[{{Branch} {et~al.}(2004{\natexlab{a}}){Branch}, {Baron}, {Thomas},
  {Kasen}, {Li}, \& {Filippenko}}]{Branch/etal:2004a}
{Branch}, D., {Baron}, E., {Thomas}, R.~C., {Kasen}, D., {Li}, W., \&
  {Filippenko}, A.~V. 2004{\natexlab{a}}, \pasp, 116, 903

\bibitem[{{Branch} {et~al.}(2003{\natexlab{b}}){Branch}, {Garnavich},
  {Matheson}, {Baron}, {Thomas}, {Hatano}, {Challis}, {Jha}, \&
  {Kirshner}}]{Branch/etal:2003}
{Branch}, D. {et~al.} 2003{\natexlab{b}}, \aj, 126, 1489

\bibitem[{{Branch} {et~al.}(2004{\natexlab{b}}){Branch}, {Thomas}, {Baron},
  {Kasen}, {Hatano}, {Nomoto}, {Filippenko}, {Li}, \&
  {Rudy}}]{Branch/etal:2004}
---. 2004{\natexlab{b}}, \apj, 606, 413

\bibitem[{{Burket} \& {Li}(2005{\natexlab{a}})}]{Burket/Li:2005a}
{Burket}, J., \& {Li}, W. 2005{\natexlab{a}}, \iaucirc, 8625, 2

\bibitem[{{Burket} \& {Li}(2005{\natexlab{b}})}]{Burket/Li:2005}
---. 2005{\natexlab{b}}, \iaucirc, 8472, 1

\bibitem[{{Chornock} {et~al.}(2006){Chornock}, {Filippenko}, {Branch}, {Foley},
  {Jha}, \& {Li}}]{Chornock/etal:2006}
{Chornock}, R., {Filippenko}, A.~V., {Branch}, D., {Foley}, R.~J., {Jha}, S.,
  \& {Li}, W. 2006, \pasp, in press (astro-ph/0603083)

\bibitem[{{Falco} {et~al.}(1999){Falco}, {Kurtz}, {Geller}, {Huchra}, {Peters},
  {Berlind}, {Mink}, {Tokarz}, \& {Elwell}}]{Falco/etal:1999a}
{Falco}, E.~E. {et~al.} 1999, \pasp, 111, 438

\bibitem[{{Ferland} \& {Persson}(1989)}]{Ferland/Persson:1989}
{Ferland}, G.~J., \& {Persson}, S.~E. 1989, \apj, 347, 656

\bibitem[{{Filippenko}(1982)}]{Filippenko:1982}
{Filippenko}, A.~V. 1982, \pasp, 94, 715

\bibitem[{{Filippenko}(1997)}]{Filippenko:1997}
---. 1997, \araa, 35, 309

\bibitem[{{Filippenko} \& {Chornock}(2003)}]{Filippenko/Chornock:2003}
{Filippenko}, A.~V., \& {Chornock}, R. 2003, \iaucirc, 8211, 2

\bibitem[{{Filippenko} {et~al.}(2003){Filippenko}, {Foley}, \&
  {Desroches}}]{Filippenko/Foley/Desroches:2003}
{Filippenko}, A.~V., {Foley}, R.~J., \& {Desroches}, L. 2003, \iaucirc, 8170, 2

\bibitem[{{Filippenko} {et~al.}(1992{\natexlab{a}}){Filippenko}, {Richmond},
  {Branch}, {Gaskell}, {Herbst}, {Ford}, {Treffers}, {Matheson}, {Ho}, {Dey},
  {Sargent}, {Small}, \& {van Breugel}}]{Filippenko/etal:1992a}
{Filippenko}, A.~V. {et~al.} 1992{\natexlab{a}}, \aj, 104, 1543

\bibitem[{{Filippenko} {et~al.}(1992{\natexlab{b}}){Filippenko}, {Richmond},
  {Matheson}, {Shields}, {Burbidge}, {Cohen}, {Dickinson}, {Malkan}, {Nelson},
  {Pietz}, {Schlegel}, {Schmeer}, {Spinrad}, {Steidel}, {Tran}, \&
  {Wren}}]{Filippenko/etal:1992}
---. 1992{\natexlab{b}}, \apjl, 384, L15

\bibitem[{{Fransson} \& {Chevalier}(1989)}]{Fransson/Chevalier:1989}
{Fransson}, C., \& {Chevalier}, R.~A. 1989, \apj, 343, 323

\bibitem[{{Gamezo} {et~al.}(2004){Gamezo}, {Khokhlov}, \&
  {Oran}}]{Gamezo/Khokhlov/Oran:2004}
{Gamezo}, V.~N., {Khokhlov}, A.~M., \& {Oran}, E.~S. 2004, Physical Review
  Letters, 92, 211102

\bibitem[{{Gamezo} {et~al.}(2005){Gamezo}, {Khokhlov}, \&
  {Oran}}]{Gamezo/Khokhlov/Oran:2005}
---. 2005, \apj, 623, 337

\bibitem[{{Gamezo} {et~al.}(2003){Gamezo}, {Khokhlov}, {Oran}, {Chtchelkanova},
  \& {Rosenberg}}]{Gamezo/etal:2003}
{Gamezo}, V.~N., {Khokhlov}, A.~M., {Oran}, E.~S., {Chtchelkanova}, A.~Y., \&
  {Rosenberg}, R.~O. 2003, Science, 299, 77

\bibitem[{{Garnavich} {et~al.}(2004){Garnavich}, {Bonanos}, {Krisciunas},
  {Jha}, {Kirshner}, {Schlegel}, {Challis}, {Macri}, {Hatano}, {Branch},
  {Bothun}, \& {Freedman}}]{Garnavich/etal:2004a}
{Garnavich}, P.~M. {et~al.} 2004, \apj, 613, 1120

\bibitem[{{G{\'o}mez} \& {L{\'o}pez}(1998)}]{Gomez/Lopez:1998}
{G{\'o}mez}, G., \& {L{\'o}pez}, R. 1998, \aj, 115, 1096

\bibitem[{{Graham} {et~al.}(2003){Graham}, {Weisz}, \&
  {Li}}]{Graham/Weisz/Li:2003}
{Graham}, J., {Weisz}, D., \& {Li}, W. 2003, \iaucirc, 8168, 1

\bibitem[{{Hamuy} {et~al.}(2000){Hamuy}, {Trager}, {Pinto}, {Phillips},
  {Schommer}, {Ivanov}, \& {Suntzeff}}]{Hamuy/etal:2000}
{Hamuy}, M., {Trager}, S.~C., {Pinto}, P.~A., {Phillips}, M.~M., {Schommer},
  R.~A., {Ivanov}, V., \& {Suntzeff}, N.~B. 2000, \aj, 120, 1479

\bibitem[{{Hillebrandt} \& {Niemeyer}(2000)}]{Hillebrandt/Niemeyer:2000}
{Hillebrandt}, W., \& {Niemeyer}, J.~C. 2000, \araa, 38, 191

\bibitem[{{Kasen} {et~al.}(2004){Kasen}, {Nugent}, {Thomas}, \&
  {Wang}}]{Kasen/etal:2004}
{Kasen}, D., {Nugent}, P., {Thomas}, R.~C., \& {Wang}, L. 2004, \apj, 610, 876

\bibitem[{{Kirshner} \& {Oke}(1975)}]{Kirshner/Oke:1975}
{Kirshner}, R.~P., \& {Oke}, J.~B. 1975, \apj, 200, 574

\bibitem[{{Koribalski} {et~al.}(2004){Koribalski}, {Staveley-Smith}, {Kilborn},
  {Ryder}, {Kraan-Korteweg}, {Ryan-Weber}, {Ekers}, {Jerjen}, {Henning},
  {Putman}, {Zwaan}, {de Blok}, {Calabretta}, {Disney}, {Minchin}, {Bhathal},
  {Boyce}, {Drinkwater}, {Freeman}, {Gibson}, {Green}, {Haynes}, {Juraszek},
  {Kesteven}, {Knezek}, {Mader}, {Marquarding}, {Meyer}, {Mould}, {Oosterloo},
  {O'Brien}, {Price}, {Sadler}, {Schr{\"o}der}, {Stewart}, {Stootman}, {Waugh},
  {Warren}, {Webster}, \& {Wright}}]{Koribalski/etal:2004}
{Koribalski}, B.~S. {et~al.} 2004, \aj, 128, 16

\bibitem[{{Kozma} {et~al.}(2005){Kozma}, {Fransson}, {Hillebrandt},
  {Travaglio}, {Sollerman}, {Reinecke}, {R{\"o}pke}, \&
  {Spyromilio}}]{Kozma/etal:2005}
{Kozma}, C., {Fransson}, C., {Hillebrandt}, W., {Travaglio}, C., {Sollerman},
  J., {Reinecke}, M., {R{\"o}pke}, F.~K., \& {Spyromilio}, J. 2005, \aap, 437,
  983

\bibitem[{{Leibundgut} {et~al.}(1993){Leibundgut}, {Kirshner}, {Phillips},
  {Wells}, {Suntzeff}, {Hamuy}, {Schommer}, {Walker}, {Gonzalez}, {Ugarte},
  {Williams}, {Williger}, {Gomez}, {Marzke}, {Schmidt}, {Whitney}, {Coldwell},
  {Peters}, {Chaffee}, {Foltz}, {Rehner}, {Siciliano}, {Barnes}, {Cheng},
  {Hintzen}, {Kim}, {Maza}, {Parker}, {Porter}, {Schmidtke}, \&
  {Sonneborn}}]{Leibundgut/etal:1993}
{Leibundgut}, B. {et~al.} 1993, \aj, 105, 301

\bibitem[{{Li} {et~al.}(2003{\natexlab{a}}){Li}, {Filippenko}, {Chornock},
  {Berger}, {Berlind}, {Calkins}, {Challis}, {Fassnacht}, {Jha}, {Kirshner},
  {Matheson}, {Sargent}, {Simcoe}, {Smith}, \& {Squires}}]{Li/etal:2003a}
{Li}, W. {et~al.} 2003{\natexlab{a}}, \pasp, 115, 453

\bibitem[{{Li} {et~al.}(2003{\natexlab{b}}){Li}, {Filippenko}, {Chornock}, \&
  {Jha}}]{Li/etal:2003}
{Li}, W., {Filippenko}, A.~V., {Chornock}, R., \& {Jha}, S. 2003{\natexlab{b}},
  \apjl, 586, L9

\bibitem[{{Livne} {et~al.}(2005){Livne}, {Asida}, \&
  {H{\"o}flich}}]{Livne/Asida/Hoflich:2005}
{Livne}, E., {Asida}, S.~M., \& {H{\"o}flich}, P. 2005, \apj, 632, 443

\bibitem[{{Matheson} {et~al.}(2000){Matheson}, {Filippenko}, {Barth}, {Ho},
  {Leonard}, {Bershady}, {Davis}, {Finley}, {Fisher}, {Gonz{\'a}lez}, {Hawley},
  {Koo}, {Li}, {Lonsdale}, {Schlegel}, {Smith}, {Spinrad}, \&
  {Wirth}}]{Matheson/etal:2000a}
{Matheson}, T. {et~al.} 2000, \aj, 120, 1487

\bibitem[{{Mazzali} {et~al.}(1998){Mazzali}, {Cappellaro}, {Danziger},
  {Turatto}, \& {Benetti}}]{Mazzali/etal:1998}
{Mazzali}, P.~A., {Cappellaro}, E., {Danziger}, I.~J., {Turatto}, M., \&
  {Benetti}, S. 1998, \apjl, 499, L49

\bibitem[{{Mazzali} {et~al.}(1997){Mazzali}, {Chugai}, {Turatto}, {Lucy},
  {Danziger}, {Cappellaro}, {della Valle}, \& {Benetti}}]{Mazzali/etal:1997}
{Mazzali}, P.~A., {Chugai}, N., {Turatto}, M., {Lucy}, L.~B., {Danziger},
  I.~J., {Cappellaro}, E., {della Valle}, M., \& {Benetti}, S. 1997, \mnras,
  284, 151

\bibitem[{{Milne} {et~al.}(2001){Milne}, {The}, \&
  {Leising}}]{Milne/The/Leising:2001}
{Milne}, P.~A., {The}, L.-S., \& {Leising}, M.~D. 2001, \apj, 559, 1019

\bibitem[{{Nugent} {et~al.}(1995){Nugent}, {Phillips}, {Baron}, {Branch}, \&
  {Hauschildt}}]{Nugent/etal:1995a}
{Nugent}, P., {Phillips}, M., {Baron}, E., {Branch}, D., \& {Hauschildt}, P.
  1995, \apjl, 455, L147

\bibitem[{{Oke} {et~al.}(1995){Oke}, {Cohen}, {Carr}, {Cromer}, {Dingizian},
  {Harris}, {Labrecque}, {Lucinio}, {Schaal}, {Epps}, \&
  {Miller}}]{Oke/etal:1995}
{Oke}, J.~B. {et~al.} 1995, \pasp, 107, 375

\bibitem[{{Phillips} {et~al.}(1992){Phillips}, {Wells}, {Suntzeff}, {Hamuy},
  {Leibundgut}, {Kirshner}, \& {Foltz}}]{Phillips/etal:1992}
{Phillips}, M.~M., {Wells}, L.~A., {Suntzeff}, N.~B., {Hamuy}, M.,
  {Leibundgut}, B., {Kirshner}, R.~P., \& {Foltz}, C.~B. 1992, \aj, 103, 1632

\bibitem[{{Pinto} \& {Eastman}(2000)}]{Pinto/Eastman:2000}
{Pinto}, P.~A., \& {Eastman}, R.~G. 2000, \apj, 530, 744

\bibitem[{{Puckett} \& {Langoussis}(2003)}]{Puckett/Langoussis:2003}
{Puckett}, T., \& {Langoussis}, A. 2003, \iaucirc, 8168, 1

\bibitem[{{Puckett} \& {Langoussis}(2005)}]{Puckett/Langoussis:2005}
---. 2005, \iaucirc, 8534, 1

\bibitem[{{Reinecke} {et~al.}(2002){Reinecke}, {Hillebrandt}, \&
  {Niemeyer}}]{Reinecke/Hillebrandt/Niemeyer:2002a}
{Reinecke}, M., {Hillebrandt}, W., \& {Niemeyer}, J.~C. 2002, \aap, 391, 1167

\bibitem[{{Ruiz-Lapuente} {et~al.}(1995){Ruiz-Lapuente}, {Kirshner},
  {Phillips}, {Challis}, {Schmidt}, {Filippenko}, \&
  {Wheeler}}]{Ruiz-Lapuente/etal:1995}
{Ruiz-Lapuente}, P., {Kirshner}, R.~P., {Phillips}, M.~M., {Challis}, P.~M.,
  {Schmidt}, B.~P., {Filippenko}, A.~V., \& {Wheeler}, J.~C. 1995, \apj, 439,
  60

\bibitem[{{Serduke} {et~al.}(2005){Serduke}, {Wong}, \&
  {Filippenko}}]{Serduke/Wong/Filippenko:2005}
{Serduke}, F.~J.~D., {Wong}, D.~S., \& {Filippenko}, A.~V. 2005, Central Bureau
  Electronic Telegrams, 269, 1

\bibitem[{{van Driel} {et~al.}(2001){van Driel}, {Marcum}, {Gallagher},
  {Wilcots}, {Guidoux}, \& {Monnier Ragaigne}}]{vanDriel/etal:2001}
{van Driel}, W., {Marcum}, P., {Gallagher}, J.~S., {Wilcots}, E., {Guidoux},
  C., \& {Monnier Ragaigne}, D. 2001, \aap, 378, 370

\bibitem[{{Verner} {et~al.}(2002){Verner}, {Gull}, {Bruhweiler}, {Johansson},
  {Ishibashi}, \& {Davidson}}]{Verner/etal:2002}
{Verner}, E.~M., {Gull}, T.~R., {Bruhweiler}, F., {Johansson}, S., {Ishibashi},
  K., \& {Davidson}, K. 2002, \apj, 581, 1154

\end{thebibliography}

\end{document}